# Superconductivity mediated by a soft phonon mode: specific heat, resistivity, thermal expansion and magnetization of YB$_6$

R. Lortz[1,‡], Y. Wang[1], U. Tutsch[1], S. Abe[1], C. Meingast[2], P. Popovich[2,3], W. Knafo[2,3], N. Shitsevalova[4], Yu. B. Paderno[4,†], and A. Junod[1].

[1] Department of Condensed Matter Physics, University of Geneva, CH-1211 Geneva 4, Switzerland.

[2] Forschungszentrum Karlsruhe, Institut für Festkörperphysik, D-76021 Karlsruhe, Germany.

[3] Physikalisches Institut, Universität Karlsruhe, D-76128 Karlsruhe, Germany.

[4] Institute for Problems of Materials Science NASU, 03680 Kiev, Ukraine.

## Abstract

The superconductor YB$_6$ has the second highest critical temperature $T_c$ among the boride family $M$B$_n$. We report measurements of the specific heat, resistivity, magnetic susceptibility and thermal expansion from 2 to 300 K, using a single crystal with $T_c$ = 7.2 K. The superconducting gap is characteristic of medium-strong coupling. The specific heat, resistivity and expansivity curves are deconvolved to yield approximations of the phonon density of states $F(\omega)$, the spectral electron-phonon scattering function $\alpha_{tr}^2 F(\omega)$ and the phonon density of states weighted by the frequency-dependent Grüneisen parameter $\gamma_G(\omega)F(\omega)$ respectively. Lattice vibrations extend to high frequencies >100 meV, but a dominant Einstein-like mode at ~8 meV, associated with the vibrations of yttrium ions in oversized boron cages, appears to provide most of the superconducting coupling and gives rise to an unusual temperature behavior of several observable quantities. A surface critical field $H_{c3}$ is also observed.

‡ Corresponding author. e-mail: Rolf.Lortz@physics.unige.ch

† Deceased on May 12, 2005

# 1. Introduction

The discovery of superconductivity at ~40 K in the metallic compound MgB$_2$ [1, 2] has stimulated a renewed interest in borides. The next highest superconducting critical temperatures in the $M$B$_n$ family are found in YB$_6$ with $T_c \leq 8.4$ K and ZrB$_{12}$ with $T_c = 6.0$ K.[3] These compounds are hard materials with a low density of states at the Fermi level. Their crystal structure, CaB$_6$-type (bcc, $Pm\overline{3}m - O_h^1$) for YB$_6$ and UB$_{12}$-type (fcc, $Fm\overline{3}m - O_h^5$) for ZrB$_{12}$, is three-dimensional and characterized by boron cages in which yttrium or zirconium atoms can develop large vibrational amplitudes. The metal-boron distance (2.76 Å in ZrB$_{12}$ and 3.01 Å in YB$_6$) is remarkably large, giving rise to unusual properties. ZrB$_{12}$ was recently investigated with respect to its specific heat, resistivity, thermal expansion and magnetic properties.[4, 5] Characteristic features of ZrB$_{12}$ were type II/1 superconductivity,[6] an enhanced gap and critical field at the surface and weak electron-phonon coupling essentially driven by a single anharmonic mode at 14 meV. In this work we turn to YB$_6$, the specific heat of which was only briefly mentioned in an early paper,[7] and find that some of these peculiarities are also present and even more dramatic. Here, a corresponding anharmonic lattice mode has softened to 8 meV, giving rise to a very unusual temperature dependence of the specific heat, resistivity and expansivity. This effect is so marked that YB$_6$ can be considered as a textbook example of superconductivity in an Einstein lattice, a limiting situation where strong-coupling theory is well assessed.[8] Another consequence of having low energy modes is that "thermal" spectroscopies, i.e. experiments sensitive to a density of energy states convolved with a thermal distribution, become efficient. In this work we show that specific heat and resistivity can indeed provide information usually taken from inelastic neutron scattering and tunneling spectroscopy, with a limited but sufficient accuracy to characterize the superconducting coupling at a quantitative level.

This article is organized as follows. In Section 2 experimental details and basic sample characterizations are given. The electronic specific heat is analyzed in Section 3. Sections 4, 5, and 6 are devoted to the deconvolution of the specific heat, resistivity, and thermal expansivity curves to obtain information on the phonon density of states $F(\omega)$, electron-phonon transport coupling function $\alpha_{tr}^2 F(\omega)$ and frequency-dependent Grüneisen parameter $\gamma_G(\omega)$ respectively. The magnetic phase diagram with four critical field lines is established in Section 7, based on different experiments. The compositional dependence is briefly addressed in Section 8, before concluding in Section 9. A wealth of experimental information is already available for YB$_6$ in the literature and we refer to Ref.[9] for a review.

# 2. Crystal growth and experimental details

The preparation of single crystals of yttrium hexaboride involved several steps: synthesis of YB$_6$ powder by borothermal reduction of Y$_2$O$_3$, compacting the powder into rods, sintering, and zone melting by inductive heating. Since the superconducting critical temperature of YB$_6$ is particularly sensitive to impurities, 5N Y$_2$O$_3$ powder and >99.5% amorphous boron were used as starting materials. The highly volatile impurities contained in the latter were eliminated during the synthesis and zone melting procedures resulting into a total impurity concentration in the crystals under study of at most 0.001% in weight.

Taking into account the peritectic melting of YB$_6$[10] and previous results on crystal growth,[11] we synthesized the initial powders with excess boron in order to lower the melting temperature. High quality single crystals were obtained for a source composition B/Y > 6.8. Other important technological parameters were the pressure of argon gas, 1.3 MPa, and the growth rate, 13 mm/h. The growth was unstable over the first few mm, yielding a two-phase mixture of YB$_4$ and YB$_6$. The process subsequently stabilized and at a definite B/Y ratio a single-phase ingot started to grow. Due to concurrent growth, one grain grew at the expense of the others and approximately <100> oriented single crystals were obtained with a length of about 30 mm.

The as-grown single crystal used for nearly all subsequent experiments was shaped into a parallelepiped bar by spark cutting and then polished using abrasive diamond paste. Its dimensions are ~12 × 3 × 1.5 mm$^3$, mass 73.5 mg, with the long axis parallel to the <100> direction and the facets perpendicular to <010> and <001>. Measurements in a magnetic field were taken with the field along the length of the sample in order to minimize the demagnetization factor ($D \cong 0.03$). The superconducting transition temperature $T_c$ was determined by four methods: resistivity (Fig. 1a), AC susceptibility (Fig. 1b), DC magnetization (Fig. 1c), and specific-heat jump at $T_c$ (see Fig. 3 below) which, in this order, are increasingly representative of the bulk volume. On average, $T_c \cong 7.2$ K, the transition width being ~2% of $T_c$ (Table I).

The DC resistivity $\rho$ was measured by a four-lead technique with current reversal, using a 5 mA current. The contacts were made with Degussa 'Leitsilber 200' conducting silver paint. The residual resistivity ratio was $\rho(300\,\text{K})/\rho(2\,\text{K}) = 3.87$. The residual resistivity $\rho(2\,\text{K}) = 9.9$ $\mu\Omega$ cm was determined in a magnetic field of 5 T to quench superconductivity; the resistivity did not vary appreciably below 9 K. No significant magnetoresistance was observed in the normal state.

The magnetization $M$ was measured in a Quantum Design MPMS-5 magnetometer, using a scan length of 4 cm. The ZFC (zero field cooled) susceptibility in the Meissner state was measured in a field of 2.7 Oe, which is ~1% of the lower critical field at $T = 0$ (Fig. 1c). The normal-state susceptibility $\chi(T)$ was obtained from the initial slope of $M(H)$ measured at 1 T intervals from 0 to 5 T. The core and Pauli contributions nearly cancel each other, resulting in a small and slightly diamagnetic susceptibility. A Curie component develops at low temperature, possibly due to traces of magnetic rare earth metals in the Y starting material (Fig. 2). The best fit is obtained by allowing a second-order term in the Pauli susceptibility:

$$\chi(T) = \chi(0) + aT^2 + \frac{C_{Curie}}{T}. \tag{1}$$

The fitted parameters are (S.I.) $\chi(0) = -9.6 \times 10^{-6}$, $a = 2.4 \times 10^{-11}$ K$^{-2}$, and $C_{\text{Curie}} = 1.8 \times 10^{-4}$ K. The Curie term is equivalent to 35 ppm at. Yb$^{3+}$.

The specific heat was measured by a generalized relaxation technique at low temperature (1.2 - 15 K)[12] and in an adiabatic, continuous-heating calorimeter at high temperature (16 - 300 K).[12] Care was taken to measure the residual field of the 14 T magnet mounted in the cryostat, zeroing this when required. Specific heat runs were taken after cooling in a field from above $T_c$ in order to achieve the best possible field penetration.

A high-resolution capacitance dilatometer[13, 14] was used to measure the thermal expansion in the temperature range 3-250 K. Data were taken both upon heating and cooling at a constant rate of 5 mK/s at low temperatures (3-15 K) and at 20 mK/s up to 250 K. Flowing He exchange gas (0-4 mbar) was used to thermally couple the sample to the dilatometer.

# 3. Electron specific heat, density of states, and coupling strength

The specific heat at low temperature in both the superconducting ($H = 0$) and normal state ($H = 1$ T) is shown in Fig. 3. The superconducting-state specific heat $C_s$ shows a sharp second-order jump at $T_c$. It vanishes at $T \rightarrow 0$ in a somewhat unusual way, since $C_s / T$ tends to a finite value $\gamma_0 = 0.03$ mJ K$^{-2}$ gat$^{-1}$ rather than zero as would be expected in a fully gapped state. This residual contribution, which may be due to an extrinsic non-superconducting fraction or to a gapless electron band, will not be discussed further as it only represents ~5% of the normal-state electronic specific heat. The normal-state specific heat is analyzed in a standard way according to the expansion

$$C_n(T \rightarrow 0) = \gamma_n T + \sum_{n=1}^{3} \beta_{2n+1} T^{2n+1} \tag{2}$$

where the first term is the electronic contribution, with $\gamma_n = \frac{1}{3}\pi^2 k_B^2 (1 + \lambda_{ep}) N(E_F)$, $k_B$ Boltzmann's constant, $\lambda_{ep}$ the electron-phonon coupling constant and $N(E_F)$ the band-structure density of states at the Fermi level including two spin directions (EDOS). The second term is the low-temperature expansion of the lattice specific heat, where $\beta_3 = \frac{12}{5} N_{Av} k_B \pi^4 \theta_D^{-3}(0)$, with $N_{Av}$ Avogadro's number and $\theta_D(0)$ the initial Debye temperature. From a fit of the normal-state data from 1.2 to 7.5 K, we obtain $\gamma_n = 0.58$ mJ K$^{-2}$ gat$^{-1}$ and $\theta_D(0) = 706$ K (369 K if according to another convention only acoustic modes are considered). One gat is $N_{Av}$ atoms, here one-seventh of a mol (Table II). The Sommerfeld constant $\gamma_n$, which is ~50% larger than that mentioned by Matthias et al.,[7] corresponds to a renormalized EDOS $(1 + \lambda_{ep}) N(E_F) = 1.73$ eV$^{-1}$cell$^{-1}$. The band structure has been calculated by several groups.[15-21] Comparing the renormalized EDOS with the recently obtained values $N(E_F) = 0.82,$[19] $0.83,$[20] and $0.93$ [21] eV$^{-1}$cell$^{-1}$, we find room for an electron-phonon renormalization factor ≈2, i.e. $\lambda_{ep} = 1.12, 1.08,$ and $0.86$, respectively. The value $\lambda_{ep} \approx 1$ is supported by independent determinations. For example, we may start from $\Delta C / \gamma_n T_c$, the normalized specific heat jump at $T_c$, as a well-defined input observable. From there we evaluate $T_c / \omega_{\ln} = 0.071$ using isotropic single-band strong-coupling formulas representing interpolated solutions of Eliashberg equations (Eq. 5.9 of Ref.[8]; $\omega_{\ln}$ is the logarithmic moment $\exp\left( \int \ln \omega \alpha^2 F(\omega) d\omega / \omega / \int \alpha^2 F(\omega) d\omega / \omega \right)$ of the Eliashberg function $\alpha^2 F(\omega)$). Assuming the conventional value $\mu^* = 0.10$ for the Coulomb pseudopotential,[22] we obtain $\lambda_{ep} = 1.01$ according to the Allen and Dynes equation.[22] A change of ±0.05 in $\mu^*$ affects the value of $\lambda_{ep}$ by ±0.15. In Section 5 we give a third independent determination $\lambda_{ep}$ =1.04 which relies on the electrical and optical conductivity and confirms the first two calculations. At this point we wish to already draw attention to the low value $\omega_{ln} = 101$ K $<< \theta_D(0) = 706$ K that results from this analysis. YB$_6$ is characterized by selective electron-phonon coupling to low-frequency modes.

The thermodynamic critical field $H_c(T)$ is obtained by numerical integration of the specific heat data:

$$-\frac{1}{2}\mu_0 V H_c^2(T) = \Delta U(T) - T\Delta S(T), \tag{3}$$

$$\Delta U(T) = \int_T^{T_c} [C_s(T') - C_n(T')]dT',$$

$$\Delta S(T) = \int_T^{T_c} \frac{C_s(T') - C_n(T')}{T'} dT',$$

where the volume $V$ and other specific quantities refer to one gram-atom. $H_c(T)$ is nearly parabolic and extrapolates to 55 mT at $T = 0$ (see Fig. 16 below). The deviation function $D(t) \equiv h - (1 - t^2)$, where $h \equiv H_c(T)/H_c(0)$ and $t \equiv T/T_c$, is a good indicator of the coupling strength.[8] The curve for YB$_6$ (Fig. 4, inset) is very similar to that of Nb$_{77}$Zr$_{23}$, for which $2\Delta(0)/k_B T_c \cong 4.2$[23] and is bounded by those of Nb and Hg, for which $2\Delta(0)/k_B T_c \cong 4.0$ and 4.6 respectively.[8] Other estimations of the gap ratio rely on the slope of the BCS semi-logarithmic plot (Fig. 4),[24] the normalized specific heat jump at $T_c$ (Eq. 4.1 of Ref.[8]) or a fit of the α-model.[25] These determinations consistently yield $2\Delta(0)/k_B T_c = 4.1 \pm 0.1$ and only differ by the weighting given to different temperature ranges. Finally we recall tunneling measurements giving $2\Delta(0)/k_B T_c = 4.02$ and $\lambda_{ep} = 0.90$ (with $\mu^* = 0$).[26] All these determinations provide evidence for medium to strong-coupling superconductivity in YB$_6$. In this respect, YB$_6$ differs from ZrB$_{12}$ which is a weakly coupled superconductor (Table II).[4, 5]

Finally, note that the shape of the electronic specific heat in the superconducting state excludes $d$-wave superconductivity. In particular, in the latter case the dimensionless ratio $\gamma_n T_c^2 / \mu_0 V H_c^2(0)$ would be nearly twice as large (3.7) as that observed (2.00±0.05).[27]

# 4. Lattice specific heat and phonon density of states

The low-temperature $T^3$ regime of the lattice specific heat does not extend beyond a few Kelvin, as shown by the large positive curvature of the normal-state curve in Fig. 3. Huge deviations with respect to the ideal Debye model persist at higher temperature (Fig. 5). The shape of the lattice specific heat in the $C/T$ plot is very uncommon, exhibiting a large low-temperature peak (Fig. 6). The specific heat at room temperature reaches only ~56% of the Dulong-Petit value, showing that the thermal energy at 300 K is still too low to excite all the spectral modes. The effective Debye temperature $\theta_{eff}(T)$ is defined as the value of $\theta$ necessary to fit the experimental specific heat at any $T$ with the equation

$$C_{ph}(T) = 9N_{Av}k_B \left(\frac{T}{\theta}\right)^3 \int_0^{\theta/T} \frac{x^4 e^x}{(e^x-1)^2}dx \ . \tag{4}$$

We have assumed $C_{ph} = C_{total} - \gamma_n T$, neglecting the anharmonic specific heat. This is justified by the estimation $C_p - C_v = (3\alpha)^2 BVT$, where $B \cong 190$ GPa is the bulk modulus and $\alpha \cong 6.1 \times 10^{-6}$ K$^{-1}$, the coefficient of linear thermal expansion (see Section 6); $(C_p - C_v)/C_p \cong 0.8\%$ at 250 K. The electronic term $\gamma_n T$ only represents ~1% of $C_{total}$ at room temperature. Starting from its initial value $\theta_{eff}(0) = \theta_D(0) = 706$ K, the effective Debye temperature passes through a deep minimum $\theta_{eff}(16\,\text{K}) = 278$ K, then increases monotonically, suggesting an asymptotic value between 1100 and 1200 K. Its room-temperature value is $\theta_{eff}(300\,\text{K}) = 1088$ K. Figure 5 shows the ideal Debye specific heat for selected values of the Debye temperature. It is clear that the data do not match any curve with a constant $\theta_D$. These plots point towards a large phonon density of states (PDOS) at energies of the order of 80 to 100 K. Analogous situations occur - albeit to a lesser extent - in Na and Al due to the presence of optical phonons[28], and in the borides ZrB$_{12}$[4] and LaB$_6$[29] for example.

The specific-heat data at high temperature are sufficiently minimally scattered to attempt a deconvolution of $C_{ph}(T)$ to extract the PDOS $F(\omega)$. More precisely, we can only obtain a substitutional spectrum, i.e. a smoothed phonon density of states $\overline{F}(\omega)$ which precisely reproduces the specific heat and low-order moments of $F(\omega)$ but may not show the true PDOS in detail. A simplified method consists of representing $F(\omega)$ by a basis of Einstein modes with constant spacing on a logarithmic frequency axis:

$$F(\omega) = \sum_k F_k \delta(\omega - \omega_k) \ . \tag{5}$$

The corresponding lattice specific heat is given by:

$$C_{ph}(T) = 3N_{Av}k_B \sum_k F_k \frac{x_k^2 e^{x_k}}{(e^{x_k}-1)^2} \tag{6}$$

where $x_k = \omega_k / T$. The weights $F_k$ are found by a least-squares fit of the lattice specific heat. The number of modes is chosen to be small enough to ensure the stability of the solution; a practical choice is $\omega_{k+1}/\omega_k = 1.75$. Note that we do not try to find the energy of each mode; we rather aim at establishing a histogram of the density in predefined frequency bins. The robustness of the fit (r.m.s deviation: <0.2% above 16 K) is demonstrated by the reproducibility of the results of two independent runs over slightly different temperature ranges (Table IV). The sum of weights exceeds the ideal value 1 by 10%; most probably part of the fitted weight in the highest energy modes in this harmonic model only serves to simulate the anharmonic contribution. Table IV also gives the generalized moments

$$\overline{\omega}_{\ln} \equiv \exp\left(\frac{\int \omega^{-1} \ln \omega F(\omega) d\omega}{\int \omega^{-1} F(\omega) d\omega}\right) \tag{7}$$

$$\left\langle \overline{\omega}^2 \right\rangle^{1/2} \equiv \left(\frac{\int \omega F(\omega) d\omega}{\int \omega^{-1} F(\omega) d\omega}\right)^{1/2}, \tag{8}$$

to be compared later with similar moments of the Eliashberg function $\alpha^2 F(\omega)$. Figure 6 illustrates the decomposition of the lattice specific heat into Einstein contributions. The PDOS obtained in this way is shown in Fig. 10 below. It consists of a background with a high cutoff frequency, as expected in view of the light and rigid boron sublattice, superposed onto a strong peak at ~8 meV which is associated with nearly free oscillations of the Y atoms in oversized boron cages. The relative weight of the latter peak, ~10% (~15% if we include both neighboring energy bins), is of the order of the fraction of Y atoms per formula unit. The question therefore arises as to what extent this low-energy region of the PDOS contributes to the electron-phonon coupling. Different answers have been given in the literature, with the main coupling being attributed to either the boron sublattice[16] or translational modes involving the yttrium ions.[26] This point is addressed in the next section, using resistivity as an experimental probe.

A comparison with standard determinations of the PDOS is instructive. YB$_6$ has been studied by inelastic incoherent neutron scattering on polycrystals.[30] According to this unpublished work, the GDOS, i.e. the generalized PDOS weighted by the scattering cross-sections of the Y and B atoms, extends to about 200 meV and exhibits a low-frequency peak at ~10 meV.[16] However, the integrated GDOS up to 15 meV only contains 1.3% of the total weight, one order of magnitude below the result from the specific heat. The GDOS and the true PDOS are expected to differ in the present case owing to the different scattering cross-sections of yttrium and boron. A determination of the PDOS by inelastic neutron scattering on single crystals is not available for YB$_6$, but dispersion curves up to ~60 meV are known for the isostructural compounds LaB$_6$ and SmB$_6$.[31, 32] In both cases, it was found that the optical modes are separated from the acoustic ones by a wide gap, which can be observed around 24 meV in the PDOS of YB$_6$ derived from the specific heat (Fig. 10). It was also pointed out that the acoustic modes of LaB$_6$ (SmB$_6$), both longitudinal and transverse, are unusually flat over the major part of the Brillouin zone, due to the non-interacting vibration of the La (Sm) ion. This gives rise to a low frequency peak at ~2.5-3 THz (~10-12 meV) in the PDOS. Although yttrium is lighter, this peak lies at ~8 meV in YB$_6$ according to the specific heat. This softening is associated with the weaker bond due to the smaller radius of the metal ion, while the size of the boron cage undergoes little change.

High frequency modes have been studied by Raman techniques in the hexaboride series (see Ref.[33] and references therein). They are associated with "internal" modes of the boron octahedra. Their energies cannot be resolved by the deconvolution of the specific heat, but their weight is included in the highest two frequency bins.

## 5. Resistivity and electron-phonon coupling

The resistivity (Fig. 7) is analyzed in a similar way. We start from the generalized Bloch-Grüneisen formula (see e.g. Ref.[34], in particular p. 212 and 219):

$$\rho_{BG}(T) = \rho(0) + \frac{4\pi m^*}{ne^2} \int_0^{\omega_{max}} \alpha_{tr}^2 F(\omega) \frac{xe^x}{(e^x - 1)^2} d\omega \qquad (9)$$

where $x \equiv \omega/T$ and $\alpha_{tr}^2 F(\omega)$ is the electron-phonon "transport coupling function". In the restricted Bloch-Grüneisen approach, one would have $\alpha_{tr}^2 F(\omega) \propto \omega^4$ and as a consequence $\rho_{BG}(T) - \rho(0) \propto T^5$, but deviations from the Debye model, complications with phonon polarizations and Umklapp processes would not justify this simplification beyond the low-temperature continuum limit, i.e. only a few Kelvin in this case. Using a decomposition into a basis of Einstein modes similar to Eq. (5),

$$\alpha_{tr}^2 F(\omega) = \frac{1}{2} \sum_k \lambda_{tr,k} \omega_k \delta(\omega - \omega_k), \qquad (10)$$

we obtain the discrete version of Eq. (9):

$$\rho_{BG}(T) = \rho(0) + \frac{2\pi}{\varepsilon_0 \Omega_p^2} \sum_k \lambda_{tr,k} \omega_k \frac{x_k e^{x_k}}{(e^{x_k} - 1)^2} \qquad (11)$$

where the fitting parameters are the dimensionless constants $\lambda_{tr,k}$. The constraint $\lambda_{tr,k} \geq 0$ is enforced. The residual resistivity $\rho(0) = 9.9$ μΩ cm is determined separately (inset of Fig. 7). The unscreened plasma frequency $\Omega_p \equiv (ne^2/\varepsilon_0 m^*)^{1/2} = 5.2$ eV is taken from extensive optical spectroscopy experiments performed on the same single crystal, to be published elsewhere.[21] The negative curvature of the resistivity at high temperature, a rather general phenomenon possibly related to the Mott limit,[35, 36] is taken into account by the empirical "parallel resistor" formula:[37]

$$\frac{1}{\rho(T)} = \frac{1}{\rho_{BG}(T) + \rho(0)} + \frac{1}{\rho_{max}}. \qquad (12)$$

The parameter $\rho_{max} = 73$ μΩ cm is fitted simultaneously to the parameters $\lambda_{tr,k}$, using data taken in either $H = 0$ ($T_c \leq T \leq 300$K) or 5 T ($2.2 \leq T \leq 300$ K). These two data sets are fitted independently (r.m.s. error ~0.2%) in order to evaluate the robustness of the results (Table IV). Only a few basis modes contribute. Two low energy modes at ~8 and ~4.5 meV clearly stand out (Fig. 10 below), in agreement with tunneling data featuring a peak in the Eliashberg function $\alpha^2 F(\omega)$ at 8.5 meV and a shoulder at ~5 meV.[26] The electron-phonon coupling parameter relevant for transport $\lambda_{tr} \equiv 2 \int \omega^{-1} \alpha_{tr}^2 F(\omega)$ is obtained from $\lambda_{tr} = \sum_k \lambda_{tr,k} = 1.04$.

Within experimental accuracy, it is equal to the electron-phonon coupling parameter relevant for superconductivity obtained in the previous section, $\lambda_{ep} \equiv 2 \int \omega^{-1} \alpha^2 F(\omega) = 1.01$. This is expected for phonon-mediated superconductors,[22, 34] but demonstrated experimentally in the present case. Finally, the value of $\omega_{\ln}$ given by

$$\omega_{\ln} \equiv \exp\left(\frac{\int \omega^{-1} \ln \omega \, \alpha_{tr}^2 F(\omega) d\omega}{\int \omega^{-1} \alpha_{tr}^2 F(\omega) d\omega}\right) = \exp\left(\frac{1}{\lambda} \sum_k \lambda_k \ln \omega_k\right) \tag{13}$$

and the alternative determination of $\omega_{\ln}$ obtained from the dimensionless specific heat jump in Section 3 both give the same value, 8.7 meV. Numerical results are summarized in Table IV, including the generalized second moment

$$\left\langle \omega^2 \right\rangle^{1/2} \equiv \left(\frac{\int \omega \, \alpha_{tr}^2 F(\omega) d\omega}{\int \omega^{-1} \alpha_{tr}^2 F(\omega) d\omega}\right)^{1/2} = \left(\frac{1}{\lambda} \sum_k \lambda_k \omega_k^2\right)^{1/2} . \tag{14}$$

When compared with thermodynamic data, the analysis of the DC and optical conductivity therefore leads to the conclusion that superconductivity is essentially driven by a single low-energy mode (or a narrow group of modes), since $\omega_{\ln}$ is very close to the low-frequency peak of the PDOS. This conclusion, which is at odds with early electronic structure and phonon modes calculations,[16] fully supports tunneling spectroscopy experiments.[26] YB$_6$ is an almost ideal case of a superconductor with an Einstein PDOS. Just as in ZrB$_{12}$,[4] most of the electron-phonon coupling arises from the large amplitude, low frequency vibrations of the loosely bound metal atoms in the oversized boron cages.

More generally, it is interesting to note that the resistivity equation (Eq. 11) can be re-expressed in a form that emphasizes the similarity with the specific heat (Eq. 6):

$$\frac{\rho_{BG}(T) - \rho(0)}{T} = \frac{R_0 k_B}{\varepsilon_0 V_p^2} \sum_k \lambda_{tr,k} E(T/T_k) \tag{15}$$

where $R_0 \equiv h/e^2 = 25.8$ K$\Omega$ is the quantum of resistance, $V_p$ the plasma frequency in Volts, and $E(x) \equiv x^2 e^x / (e^x - 1)^2$ the normalized Einstein specific heat at a temperature $T$ due to a mode with a characteristic temperature $T_k$. In the case of coupling to a single mode, this reduces to

$$\frac{\rho_{BG}(T) - \rho(0)}{C_{ph} T} = \frac{R_0 k_B}{\varepsilon_0} \frac{\lambda_{tr}}{V_p^2} = \text{constant} \tag{16}$$

This scaling of $C_{ph}$ and $[\rho_{BG}(T) - \rho(0)]/T$ is approximately obeyed in ZrB$_{12}$ and YB$_6$ due to the predominance of one soft mode. Grimvall[34] noted that a similar relation holds in the high temperature limit for any number of modes. As no large variation is expected in the plasma frequency $\Omega_p \equiv (e^2/\varepsilon_0 m_e V_{cell})^{1/2}$ of trivalent hexaborides which have one free carrier per unit cell,[9] Eq. (16) immediately shows that $\lambda_{tr}$ in LaB$_6$, which has a room-temperature phonon resistivity $\rho(300) - \rho(0) = 8.9$ $\mu\Omega$ cm,[38] is much weaker than $\lambda_{tr}$ in YB$_6$, for which $\rho(300) - \rho(0) = 28.4$ $\mu\Omega$ cm. Indeed superconductivity has been reported to occur only below 0.1 K in LaB$_6$. More information on the possibilities and limitations of the deconvolution of the resistivity to obtain the $\alpha_{tr}^2 F(\omega)$ function may be found in the paper of Igalson *et al.*[39]

## 6. Thermal expansivity and anharmonicity

Thermal-expansion experiments were undertaken to give three types of information: (i) confirmation of the main features of the PDOS; (ii) evaluation of the volume dependence of phonon modes and electronic density of states; (iii) determination of the variation of $T_c$ with pressure. The linear thermal expansivity $\alpha(T)$ for a cubic system is given by:

$$\alpha(T) \equiv \frac{1}{L}\left(\frac{\partial L}{\partial T}\right)_p = \frac{\kappa_T}{3}\left(\frac{\partial S}{\partial V}\right)_T , \tag{17}$$

where $\kappa_T$ is the isothermal compressibility. The expansivity is closely related to the specific heat at constant volume via the Grüneisen parameters (see e.g. Ref. [34]):

$$\alpha(T) = \frac{\kappa_T}{3V}\left(\gamma_{G,el}C_{el} + \gamma_{G,ph}C_{ph}\right) \tag{18}$$

where the electronic Grüneisen parameter $\gamma_{G,el} = \partial \ln \gamma_n / \partial \ln V$ gives a measure of the volume dependence of the Sommerfeld constant and the phonon Grüneisen parameter $\gamma_{G,ph} \equiv -\partial \ln \omega / \partial \ln V$ represents the anharmonicity of the lattice vibrations. In this simple form, we can make use of the known components $C_{el}(T)$ and $C_{ph}(T)$ of the specific heat in the normal state and adjust $\gamma_{G,el}$ and $\gamma_{G,ph}$ to fit the normal-state expansivity curve $\alpha(T)$ at low temperature. As in the case of the specific heat, a plot of $\alpha / T$ versus $T^2$ is most suitable for displaying the results. The fitted parameters $\gamma_{G,el} = 15\pm3$ and $\gamma_{G,ph} = 9\pm1$ are stable when the upper limit of the fit is varied between 50 and 120 K$^2$. This determination of $\gamma_{G,ph}$ is representative of the lowest frequencies of the phonon spectrum; at higher temperatures the quality of the fit degrades rapidly. With $\gamma_{G,el}$ we determine the electronic component of the expansivity, $\alpha_{el}(T)/T = (2.6\pm0.5) \times 10^{-9}$ K$^{-2}$. In these evaluations the bulk modulus $\kappa_T^{-1} = 190$ GPa has been estimated from Fig. 2 of Ref.[33]; the value given by band-structure calculations is $\kappa_T^{-1} = 179$ GPa.[20]

At higher temperature, the frequency dependence of the phonon Grüneisen parameter must be taken into account. Modes which are characterized by a large $\gamma_{G,ph}(\omega)$ are more heavily weighted in the thermal expansion than in the lattice specific heat. This is exemplified by the expansivity data shown in Fig. 8 over the full temperature range, to be compared with the specific heat in Fig. 5. The broad anomaly which appears around 50 K in Fig. 8 is evidence for a large volume dependence in some low-frequency modes. In order to evaluate the energy of these modes, we fit the phonon expansivity over the full temperature range in a similar manner to the resistivity and the specific heat, using the same set of Einstein frequencies. Equation (19) below, similar to Eq. (6) and (11), allows the parameters $\gamma_{G,k}F_k$ to be extracted for each frequency $\omega_k$ :

$$\alpha_{ph}(T) = \alpha(T) - \alpha_{el}(T) = \frac{N_{Av}k_B\kappa_T}{V}\sum_k \gamma_{G,k}F_k \frac{x_k^2 e^{x_k}}{(e^{x_k}-1)^2}. \tag{19}$$

The PDOS weighted by the frequency-dependent Grüneisen parameter, $\gamma_{G,ph}(\omega)F(\omega)$, is represented in Fig. 10 together with other spectra. The 8-meV and 4.5-meV modes are heavily weighted with $\gamma_{G,k} \cong 7$ and 9 respectively, whereas the other modes are much less anharmonic with $\gamma_{G,k}$ values below 2.[40] Similarly to MgB$_2$ and ZrB$_{12}$, the modes which give rise to a large electron-phonon coupling are anharmonic.

The pressure dependence of $T_c$ is obtained from the Ehrenfest relation

$$\Delta\alpha = \frac{1}{3V}\frac{\Delta C}{T_c}\left(\frac{\partial T_c}{\partial p}\right)_T, \tag{20}$$

where $\Delta\alpha$ and $\Delta C$ represent discontinuities of $\alpha$ and $C$ at the second-order transition. The experimentally determined step $\Delta\alpha = -(3.5\pm0.5)\times10^{-8}$ K$^{-1}$ (Fig. 9) corresponds to $-0.53\pm0.08$ K/GPa for the initial pressure dependence of $T_c$. Again assuming $\kappa_T^{-1} = 190$ GPa, one obtains the fractional volume dependence of the critical temperature $\partial \ln T_c / \partial \ln V = 14\pm2$.

The fractional volume dependences of the critical temperature and Sommerfeld constant are unusually large, 14 and 15 respectively. The fact that they are nearly equal is associated with the fact that the relative jumps of the expansivity and specific heat are equal in magnitude but opposite in sign. This follows from Eq. (18) and (20):

$$\left(\frac{\partial \ln T_c}{\partial \ln V}\right)_T = -\frac{\Delta\alpha / \alpha_{el}(T_c)}{\Delta C / C_{el}(T_c)}\left(\frac{\partial \ln \gamma_n}{\partial \ln V}\right)_T \tag{21}$$

The roles of the phonon modes and the EDOS in the volume dependence still have to be elucidated. If the transition temperature is derived from a McMillan-type relation,[41] by neglecting the volume dependence of the screened Coulomb repulsion parameter $\mu^*$ and recalling that the Eliashberg function is strongly peaked, $\alpha^2 F(\omega) \approx \lambda\omega_E\delta(\omega - \omega_E)$ we obtain[34]

$$\frac{d \ln T_c}{d \ln V} = -\gamma_{G,ph}(\omega_E) + f(\lambda_{ep},\mu^*)\frac{d \ln \lambda_{ep}}{d \ln V} \tag{22}$$

where $f(\lambda_{ep},\mu^*)$ is easily calculated from McMillan's equation and takes the value ~1.5 in the present case with $\lambda_{ep} \cong 1$, $\mu^* \cong 0.1$. A second equation describes the fractional volume dependence of $\gamma_n \propto N(E_F)(1+\lambda_{ep})$:

$$\frac{d \ln \gamma_n}{d \ln V} = \frac{d \ln N(E_F)}{d \ln V} + \frac{\lambda_{ep}}{1+\lambda_{ep}}\frac{d \ln \lambda_{ep}}{d \ln V} \tag{23}$$

From Eq. (22) and (23), $\partial \ln N(E_F)/\partial \ln V = 8.2 \pm 4$ and $\partial \ln \lambda_{ep}/\partial \ln V = 13.7 \pm 2$. The variation of $\gamma_n$ with the volume is therefore due to the variation of both the EDOS and the phonon-dependent renormalization in an approximately equal ratio.[42]

In a final step we may write $\lambda_{ep} = \eta/M\omega_E^2$, where $\eta$ is the Hopfield electronic parameter. It follows that

$$\frac{d \ln \lambda_{ep}}{d \ln V} = \frac{d \ln \eta}{d \ln V} + 2\gamma_{G,ph}(\omega_E) . \tag{24}$$

The first term of the right-hand side is found to be small, $d \ln \eta / d \ln V = -0.3 \pm 3$, compared to the second one, $2\gamma_{G,ph}(\omega_E) = 14 \pm 2$. Therefore we may say that the large and positive values of $\partial \ln \lambda_{ep}/\partial \ln V$ and $\partial \ln T_c/\partial \ln V$ are essentially due to the anharmonicity of the 8-meV mode. When the volume of the boron cages increases, the force constant which determines the frequency of the strongly coupled soft mode decreases and the latter moves closer to the favorable region $\approx 6k_B T_c$ where the functional derivative $\delta T_c/\delta\alpha^2 F(\omega)$ is largest.[8]

## 7. Magnetic phase diagram

The superconducting phase diagram in the $H$-$T$ plane was investigated by magnetoresistance, magnetization and specific heat measurements, allowing the critical fields $H_{c1}(T)$, $H_c(T)$, $H_{c2}(T)$ and $H_{c3}(T)$ to be determined. The resistive transitions in fields 0 to 0.5 T are shown in Fig. 11. They are measured with both the field and the current parallel to the long axis of the crystal. The extrapolation of the steepest part of the transition to $R = 0$ is used to define the surface critical field $H_{c3}(T)$, which is well separated from $H_{c2}(T)$ as already observed in the dodecaboride ZrB$_{12}$.[5]

The DC magnetization at temperatures from 2 to 6 K is shown in Fig. 12. Our data agree with those of Kunii *et al.*[43] The shape is typical of a type-II superconductor. The sharp minimum at the border of the Meissner and mixed-state regions defines $H_{c1}(T)$ and the break in the slope between the mixed-state and normal-state regions defines $H_{c2}(T)$. It is remarkable that no anomaly can be detected in the magnetization at $H_{c3}(T)$ where the resistance vanishes (Fig. 13), thus confirming the superficial nature of the third critical field. No anomaly is detected in the specific heat either at $H_{c3}(T)$. Note that the scale of Fig. 13 is enlarged by a factor of 5000 with respect to that of Fig. 12. Although on this scale the transition at $H_{c2}$ appears to be rounded, $H_{c2}$ nevertheless remains well defined using an extrapolation of the linear parts from above and below the transition, as expected in the Ginzburg-Landau regime (Fig. 13, inset).

Finally the specific heat was measured in several fields from 0 to 1 T. These measurements give an independent bulk determination of $H_{c2}(T)$. The criterion used here is the midpoint of the step at the transition (Fig. 14). No anomaly is seen at $H_{c1}(T)$, thus establishing that YB$_6$ is a type-II/2 superconductor, unlike ZrB$_{12}$ which was of type-II/1[5] Note that in Fig. 14 only the electronic part $C_e/T$ of the total specific heat is shown. The rise of $C_e(H)/T$ with respect to the zero-field curve $C_e(0)/T$, shown in Fig. 15, is due to the contribution of quasi-normal vortex cores and the excitations of the vortex lattice. The trends observed in the data suggest that the model of Caroli *et al.*,[44, 45] i.e. $C_e(H)/T = \gamma_n H/H_{c2}(0)$, would be obeyed at temperatures below $\sim 0.1\,T_c$.

The above information is summarized in the phase diagram, Fig. 16. The Maki parameter $\kappa_1(T) \equiv 2^{-1/2} H_{c2}(T)/H_c(T)$ shown in the inset extrapolates to $\kappa_1(T_c) \equiv \kappa \cong 3$ and increases at low temperature as usual. Note that $\kappa$ is too low for the approximate Ginzburg-Landau relation $H_{c1}(T) \cong 2^{-1/2} H_c(T)\ln\kappa_1/\kappa_1$ to hold. Instead we use $\kappa_1(T)$ to recalculate $H_{c1}(T)$ according to the numerical work of Harden and Arp[46] (lower dashed line in Fig. 16). These recalculated values nevertheless underestimate $H_{c1}(T)$ taken from magnetization curves by 10 to 20%.

The critical field $H_{c3}(T)$, which describes the persistence of superconductivity over a layer of thickness comparable to the coherence length $\xi$ when the field is applied parallel to a flat surface, should ideally be $1.695\,H_{c2}(T)$.[47, 48] Indeed, at 2.4 K (the lowest temperature at which $H_{c3}(T)$ could be measured), $H_{c3}(T)/H_{c2}(T) = 1.66$. However, while $H_{c2}(T)$ has a negative curvature which fits well with the WHH theory,[49] $H_{c3}(T)$ has a small but positive curvature. Using a polynomial fit, $H_{c3}(T)$ extrapolates to $\sim 2.2\,H_{c2}(0)$ at $T = 0$ (Fig. 16). A qualitatively similar behavior was found in ZrB$_{12}$.[5] This exceedingly large ratio may be explained here by a decrease in the mean free path with respect to the bulk over a thin layer at the surface. An increase of the gap at the surface with respect to the bulk volume (as found in ZrB$_{12}$[4, 5, 50, 51]) is excluded in YB$_6$, since in this case tunneling and specific heat determinations agree on the gap value.

# 8. Compositional dependence

A broad range of superconducting critical temperatures $1.5 \leq T_c \leq 8.4$ K has been reported in the literature depending on the conditions of preparation.[9] In our preliminary work, $T_c$ onsets from 6.5 to 7.6 K were obtained. EDX (energy-dispersive X-ray diffraction) revealed a correlation between $T_c$ and the Y/B ratio, the higher $T_c$ corresponding to a smaller boron concentration ($T_c(\text{YB}_{6+x}) \cong 6.25 - 4.3x \pm 0.25$ K). Literature data confirm this tendency.[9] The lattice constant was found to be almost insensitive to the Y/B ratio, changing from 4.1002(5) Å for YB$_{5.7}$ with $T_c = 7.6$ K to 4.1000(5) Å for YB$_{5.9}$ with $T_c = 6.6$ K. These values are in agreement with published data.[9] The possible influence of strains on $T_c$ cannot be excluded: the high value $T_c = 8.4$ K reported by Fisk *et al.*[3] was obtained by splat-cooling of arc-melted samples with nominal composition YB$_6$. We tried to 'anneal out' residual strains in different samples by a 36 hr heat treatment at 1600 C in 100 bar argon pressure, followed by slow cooling. The $T_c$ onset of low-$T_c$ samples (6.5 K) did not change, that of high $T_c$ samples (7.6-7.8 K) decreased by about half a Kelvin. In all cases the main effect was a broadening of the transition by 2 to 3 K. While such experiments tend to support the idea that strains increase $T_c$, the effect of losses during the heat treatment cannot be excluded.

The highest $T_c$ is obtained for a B/Y ratio below 6. In this sense the reference crystal with $T_c$ = 7.2 K studied in the previous sections is not ideal because of the presence of boron vacancies. Therefore we briefly studied two other crystals with $T_c^{onset}$ = 6.5 K and 7.6 K. Their resistivity was only found to differ from that of the reference sample by the residual term, while the temperature-dependent part remained unchanged (Table V). Magnetization curves were measured with increasing field and integrated to get the thermodynamic critical field $H_c(T)$ (Fig. 17, Table V). Irreversibility may introduce some error by delaying the entry of vortices, leading to an overestimation of $H_c(T)$. However, we found that the magnetic determination of $H_c(T)$ was only 4% above that obtained from the specific heat for the reference sample, showing that the branch with increasing field is close to equilibrium. The net result is that $T_c$ is positively correlated with $H_c(0)$, $\gamma_n$ and therefore the EDOS, as is the case for most BCS superconductors. The change in the EDOS is confirmed by the DC magnetic susceptibility. The curve for the sample with $T_c$ = 6.5 K is similar to that shown in Fig. 2 and described by the parameters $\chi(0) = -16.8 \times 10^{-6}$, $a = 4.0 \times 10^{-11}$ K$^{-2}$, and $C_{Curie}$ = $6.5 \times 10^{-4}$ K (see Section 2). Compared to the reference sample, the susceptibility is more diamagnetic, which is attributed to a smaller Pauli contribution. The Curie term is equivalent to 120 ppm Yb$^{3+}$.

# 9. Conclusion

Specific-heat, resistivity and thermal-expansion experiments performed on high quality single crystals have been used to characterize YB$_6$. This superconductor has a low density of states at the Fermi level. Some sample dependence of $T_c$ can be traced to the variation of the B/Y ratio, which in turn influences the EDOS. The specific heat in the superconducting state is typical of a single-band, isotropic BCS superconductor; however, the electron-phonon interaction is much stronger than for the other superconducting borides, in particular ZrB$_{12}$ and MgB$_2$. A common feature of these borides is non-uniform coupling to selected phonon modes. Whereas the strongly coupled modes lie at high energy ~60 meV for MgB$_2$,[52] they lie at low energy ~15 meV for ZrB$_{12}$[4] and ~8 meV for YB$_6$, which partly explains their relatively low $T_c$. Similarly to LaB$_6$,[29] these low frequency modes are associated with the vibration of Y or Zr atoms loosely bound in oversized boron cages. The reason for the lower characteristic frequency in YB$_6$ compared to ZrB$_{12}$ is neither to be found in the mass of the metal ions, nor in the coordination number which is 24 in both cases. It is rather to be associated with the longer metal to boron bond, 3.03 Å in YB$_6$[53] instead of 2.76 Å in ZrB$_{12}$.[54] This longer distance leads to a weaker force constant and larger vibrational amplitude, which in turn favors superconductivity. The thermal expansion indeed indicates that $T_c$ will decrease with pressure and that these modes are strongly anharmonic. As for magnetic properties, YB$_6$ is a type-II superconductor with $\kappa \cong 3$ and clearly shows a third upper critical field at the surface, a rather rare occurrence.

Owing to the low characteristic energy of the phonons which mediate superconducting pairing, YB$_6$ has been found to be an almost ideal system where on one hand specific heat can substitute for inelastic neutron scattering (which is plagued by the absorption of [10]B) to study the PDOS, and on the other hand resistivity can be substituted for tunnelling to study the electron-phonon coupling function. The spectral resolution of these procedures is limited and only makes sense on a logarithmic energy scale; nevertheless this technique is found to be reproducible and able to give remarkably consistent values of $\lambda_{ep}$, allowing significant comparisons to be made between borides.

## Acknowledgements

Stimulating discussions with J. Teyssier and J. Geerk are gratefully acknowledged. We thank E. Giannini, P. Lezza, B. Revaz, and A. Naula for their help in XRD, EDX and metallurgy. This work was supported by the National Science Foundation through the National Centre of Competence in Research "Materials with Novel Electronic Properties–MaNEP", and INTAS Project 03-51-3036.

## Tables

Table I. Critical temperature and transition width of the YB$_6$ crystal measured by different methods.

|  | $T_c$ midpoint (K) | $\Delta T_c$ (K) |
|---|---|---|
| Resistivity in zero field @ 1 mA/mm$^2$ | 7.20 | <0.2 (0-100%) |
| AC susceptibility @ 8 kHz, 0.01 Oe | 7.24 | 0.15 (10-90%) |
| Meissner magnetization @ 2.7 Oe | 7.13 | 0.20 (10-90%) |
| Specific-heat jump in zero field | 7.15 | 0.13 (10-90%) |

Table II. Characteristic parameters of YB$_6$ compared to ZrB$_{12}$. $T_c$, superconducting transition temperature; $RRR$, residual resistivity ratio; $V$ and $M$, mean atomic volume and mass, respectively; $a$, lattice constant; $\gamma_n$, coefficient of the linear term of the normal-state specific heat at $T \rightarrow 0$; $\Delta C / T_c$, specific-heat jump at $T_c$; $\Delta C / \gamma_n T_c$, normalized specific-heat jump; $\Theta_D(0)$, initial Debye temperature; $S(300)$, total entropy at 300 K; $E_c$, condensation energy at $T \rightarrow 0$; $2\Delta(0)/k_B T_c$, normalized superconducting gap; $\chi(0)$, normal-state magnetic susceptibility at $T \rightarrow 0$; $N_{sb}(E_F)$, bare density of states at the Fermi level (per 7-atom cell for YB$_6$,[19-21] per 13-atom quarter-cell for ZrB$_{12}$.[21, 55]); $1 + \lambda_{ep}$, electron-phonon renormalization factor as determined from $\gamma_n$ and $N_{sb}(E_F)$; $\lambda_{ep}$, electron-phonon coupling constant as determined from $T_c$ and $\omega_{ln}$.[22]

|  | YB$_6$ | ZrB$_{12}$ |
|---|---|---|
| $T_c$ (K) | $7.15 \pm 0.05$ | $5.96 \pm 0.05$ |
| $RRR$ | $3.87 \pm 0.03$ | $9.33 \pm 0.03$ |
| $V$ (cm$^3$ gat$^{-1}$) | 5.929 | 4.68 |
| $M$ (g gat$^{-1}$) | 21.97 | 17.0 |
| $a$ (nm) | $0.41002 \pm 0.00005$ | 0.7407 |
| $\gamma_n$ (mJ K$^{-2}$ gat$^{-1}$) | $0.58 \pm 0.02$ | $0.34 \pm 0.02$ |
| $\Delta C / T_c$ (mJ K$^{-2}$ gat$^{-1}$) | $1.18 \pm 0.02$ | $0.56 \pm 0.02$ |
| $\Delta C / \gamma_n T_c$ | $2.02 \pm 0.1$ | $1.66 \pm 0.1$ |
| $\theta_D(0)$ (K) | $706 \pm 20$ | $970 \pm 20$ |
| $S(300)$ (J K$^{-1}$ gat$^{-1}$) | $13.5 \pm 0.1$ | $9.3 \pm 0.1$ |
| $E_c$ (mJ/gat) | $7.15 \pm 0.2$ | $3.2 \pm 0.1$ |
| $2\Delta(0)/k_B T_c$ | $4.1 \pm 0.1$ | $3.7 \pm 0.1$ |
| $\chi(0)$ (S.I.) | $-9.6 \times 10^{-6}$ | $-20.8 \times 10^{-6}$ |
| $N_{sb}(E_F)$ (eV cell)$^{-1}$ | 0.81-0.93 | 1.57-1.59 |
| $1 + \lambda_{ep}$ (from $\gamma_n$) | 1.86-2.14 | 1.18-1.19 |
| $\lambda_{ep}$ (from $T_c$) | 1.01 | 0.61-0.65 |

Table III. $H_c(0)$, thermodynamic critical field at $T \to 0$ obtained from specific heat (C) and magnetization (M) measurements; $(dH_c / dT)_{Tc}$, slope of the thermodynamic critical field at $T \to T_c$; $H_{c1}(0)$, lower critical field at $T \to 0$; $H_{c2}(0)$, upper critical field at $T \to 0$; $(dH_{c2} / dT)_{Tc}$, slope of the upper critical field at $T \to T_c$; $H_{c3}(0)$, surface upper critical field at $T \to 0$; $\kappa \equiv \kappa_1(T_c)$, Maki parameter.

| | |
|---|---|
| $H_c(0)$ (mT) | 55 (C), 58 (M) |
| $(dH_c / dT)_{Tc}$ (mT/K) | -15.7 |
| $H_{c1}(0)$ (mT) | $20 \sim 25$ |
| $H_{c2}(0)$ (mT) | 295 |
| $(dH_{c2} / dT)_{Tc}$ (mT/K) | -59 |
| $H_{c3}(0)$ (mT) | ~640 |
| $\kappa_1(T_c)$ | 3.0 |

Table IV. Results of the fit of the specific heat in terms of $F(\omega)$, and the fit of the resistivity in terms of $\alpha^2 F_{tr}(\omega)$. $\omega_k$, energy of the modes in a geometrical series $\omega_{k+1} = 1.75\omega_k$. The origin is arbitrarily chosen. $F_k$, weight associated with each mode in $F(\omega)$; two sets of numbers show the results of two independent runs. A (-) sign means that this mode was not included in the fit. $\lambda_{tr,k} \equiv 2(\alpha^2 F)_{tr,k} / \omega_k$, partial contribution of the mode to $\lambda_{tr}$; two sets of numbers show the results of two independent runs. $\overline{\omega}_{\ln}$, $\langle \overline{\omega}^2 \rangle^{1/2}$: generalized moments of $F(\omega)$; $\omega_{\ln}$, $\langle \omega^2 \rangle^{1/2}$: generalized moments of $\alpha^2 F_{tr}(\omega)$ (see text).

| $\omega_k$ (meV) | $F_k / \omega_k$ (eV$^{-1}$) 1$^{st}$ run, 2-300 K, $H = 1$ T (2$^{nd}$ run, 16-300 K, $H = 0$) | $\lambda_{tr,k} = (\alpha^2 F)_k / \omega_k$ 1$^{st}$ run, 2-300 K, $H = 5$ T (2$^{nd}$ run, 7.5-300 K, $H = 0$) |
|---|---|---|
| 127 | 3.37 (3.40) | 0 (0) |
| 72.7 | 4.60 (4.61) | 0.101 (0.117) |
| 41.6 | 4.07 (4.01) | 0 (0) |
| 23.8 | 0.80 (0.93) | 0 (0) |
| 13.6 | 3.37 (3.21) | 0 (0.006) |
| 7.76 | 13.0 (13.1) | 0.724 (0.743) |
| 4.43 | 0.83 (0.93) | 0.180 (0.186) |
| 2.53 | 0.12 (-) | 0.024 (0) |
| 1.45 | 0.009 (-) | (-) |
| 0.83 | 0.013 (-) | (-) |
| | $\sum F_k = 1.103$ (1.105) | $\lambda_{tr} = \sum \lambda_{tr,k} = 1.03$ (1.05) |
| | $\overline{\omega}_{\ln} = 20.0$ (20.2) meV | $\omega_{\ln} = 8.5$ (9.0) meV |
| | $\langle \overline{\omega}^2 \rangle^{1/2} = 53.9$ (54.0) meV | $\langle \omega^2 \rangle^{1/2} = 23.8$ (25.2) meV |

Table V. Parameters of the fit $H_c(T)/H_c(0) = 1 - (T/T_c)^2$ of the magnetization data shown in Fig. 17 for three YB$_{6+x}$ crystals: $T_c$, critical temperature; $H_c(0)$, thermodynamic critical field at zero temperature. $\gamma_n$, Sommerfeld constant estimated assuming $\gamma_n T_c^2 / H_c^2(0) =$ constant; $RRR$, residual resistivity ratio. The resistance of the sample with $T_c = 7.4$ K (temperature at which $H_c(T) \rightarrow 0$) vanishes at 7.6 K.

| $T_c$ (K) | 6.5 | 7.2 | 7.4 |
|---|---|---|---|
| $H_c(0)$ (mT) | 48 | 58 | 61 |
| $\gamma_n$ (mJ K$^{-2}$ gat$^{-1}$) | 0.50 | 0.58 (reference) | 0.62 |
| $RRR$ | 3.05 | 3.87 | 4.58 |

# Figure captions

Fig. 1.  Superconducting transition of the YB$_6$ crystal observed by (a) resistivity, (b) AC susceptibility (8 kHz, 0.01 G r.m.s.), and (c) Meissner susceptibility (field cooled, 2.7 G).

Fig. 2.  Magnetic susceptibility of YB$_6$ in the normal state as a function of the temperature. Full line: fit (see text). Dashed line: non-Curie part of the fit.

Fig. 3.  Total specific heat $C/T$ of YB$_6$ in the superconducting state in zero field (closed symbols) and in the normal state in 1 Tesla (open symbols) versus the temperature squared.

Fig. 4.  BCS plot of the electronic specific heat in the superconducting state normalized by the Sommerfeld constant $\gamma_n$, versus the inverse reduced temperature, data (symbols) and BCS weak-coupling limit (line). The residual contribution $\gamma_0 T$ has been subtracted. Inset: deviation function of the thermodynamic critical field (symbols) and BCS weak-coupling limit (dashed line).

Fig. 5.  Lattice specific heat of YB$_6$ versus the temperature up to room temperature. The long-dashed line shows the Debye specific heat calculated using a constant Debye temperature equal to its minimum $\theta_D(16\,\text{K}) = 278$ K, the short-dashed line using the effective value at room temperature $\theta_D(300\,\text{K}) = 1088$ K. Inset: effective Debye temperature versus temperature.

Fig. 6.  Lattice specific heat divided by the temperature versus the temperature showing the decomposition into Einstein terms. The labels $k$ correspond to Einstein temperatures $\Theta_{E,k} = 90\,\text{K} \cdot 1.75^k$, i.e. (from left to right) 51, 90, 158, 276, 482, 844, and 1477 K.

Fig. 7.  Normal-state resistivity of YB$_6$ versus the temperature. Dashed line: residual resistivity. Crosses: residuals of the fit in %. Inset: expanded low-temperature data and polynomial fit. Superconductivity is quenched by a field of 1 T.

Fig. 8.  Linear thermal expansivity of YB$_6$ versus temperature. Inset: Grüneisen parameter versus temperature, assuming a bulk modulus $\kappa_T^{-1} = 190$ GPa.

Fig. 9. Difference between the normal-state and superconducting-state linear thermal expansivity near $T_c$. The idealized jump is shown by a dotted line. Inset: expansivity in the normal and in the superconducting state.

Fig. 10. Phonon density of states $F(\omega)$ deconvoluted from the specific heat, electron-phonon transport coupling function $\alpha_{tr}^2 F(\omega)$ deconvolved from the resistivity, and spectral anharmonicity function $\gamma_G(\omega)F(\omega) \equiv -(\partial \ln \omega / \partial \ln V)F(\omega)$ deconvoluted from the thermal expansion. Fits are performed with $\delta$-functions $F_k\delta(\omega - \omega_k)$, $(\alpha_{tr}^2 F)_k\delta(\omega - \omega_k)$, and $(\gamma_G F)_k\delta(\omega - \omega_k)$, respectively, on a basis of Einstein frequencies $\omega_{k+1} = 1.75\omega_k$ (see Fig. 6). In order to reflect the spectral density, the $\delta$-functions of the PDOS are represented by rectangles having a width $\Delta\omega_k \equiv 1.75^{1/2}\omega_k - \omega_k / 1.75^{1/2} \cong 0.57\omega_k$ and a height $F_k / \Delta\omega_k$. In a similar way, the $\delta$-functions of the $\alpha_{tr}^2 F(\omega)$ function are represented by closed circles at a height $(\alpha_{tr}^2 F)_k / \Delta\omega_k \cong 0.88\lambda_k$, and those of $\gamma_G(\omega)F(\omega)$ by triangles at a height $(\gamma_G F)_k / \Delta\omega_k$. The dashed lines are guides to the eye.

Fig. 11. Resistivity of YB$_6$ near $T_c$ as a function of the temperature in different magnetic fields.

Fig. 12. Magnetization of YB$_6$ as a function of the magnetic field for different temperatures, virgin curves with increasing field ($M$ in emu/cm$^3$; $4\pi M$ in G). Inset: sample hysteresis curve at 4 K (similar curves are obtained from 2 to 6 K).

Fig. 13. Detail of the previous plot at $T = 5$ K, expanded 100 times in the inset and 5000 times in the main frame. The normal-state magnetization $M_n$ has been subtracted. Note the absence of any measurable diamagnetism at $H_{c3} = 1700$ G where the resistance drops to zero. Fluctuation diamagnetism sets in smoothly near 1500 G. The upper critical field $H_{c2} = 1270$ G defined by the intersection of the extrapolated linear sections in the inset coincides with the position of the specific heat jump.

Fig. 14. Electronic specific heat of YB$_6$ divided by the temperature versus the temperature for different magnetic fields.

Fig. 15. Specific heat difference $[C(H,T) - C(0,T)] / T$ normalized by the Sommerfeld constant versus $T / T_c$ in different magnetic fields. The value calculated within the Caroli-Matricon-de Gennes approximation is shown by full lines at $T \to 0$.

Fig. 16. Phase diagram of YB$_6$ in the $H - T$ plane. From top to bottom: third critical field $H_{c3}(T)$ defined by zero resistance for $H$ parallel to the current and the surface; second upper critical field $H_{c2}(T)$ given by the position of the specific heat jump (closed diamonds) and the knee of the magnetization (open diamonds); thermodynamic critical field $H_c(T)$ obtained by integration of the specific heat $C/T$ (open circles); lower critical field $H_{c1}(T)$ given by the position of the sharp minimum of the magnetization (closed circles). All lines are polynomial fits to the data, except for the $H_{c1}(T)$ line, which is recalculated based on $H_c(T)$ and the Maki parameter $\kappa_1(T)$ (see text). Inset: variation of $\kappa_1$ with the temperature.

Fig. 17. Thermodynamic critical field $H_c(T)$ of boron-rich (lower $T_c$) and boron-deficient (higher $T_c$) samples. In this plot, $H_c(T)$ is obtained by integration of the magnetization curves.

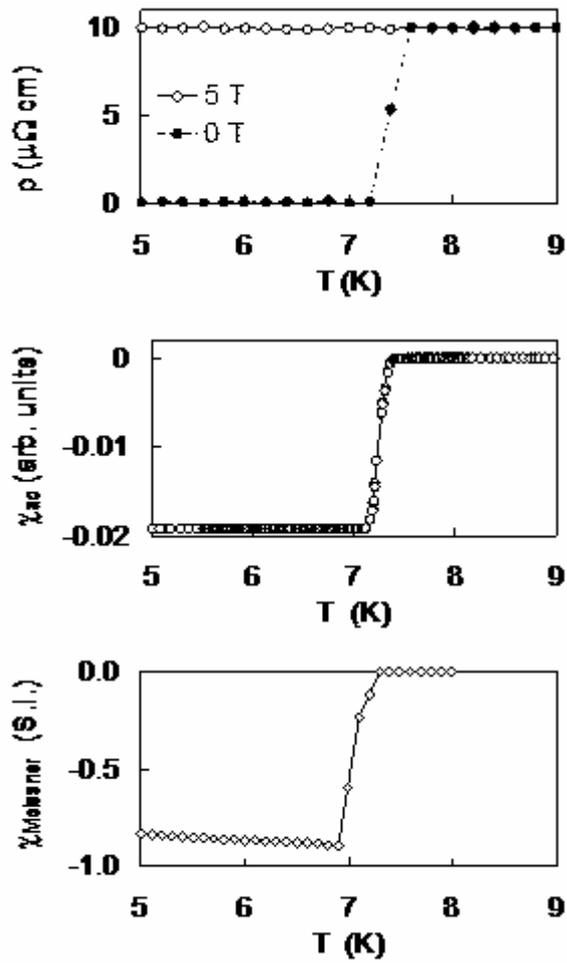

**1a 1b 1c**

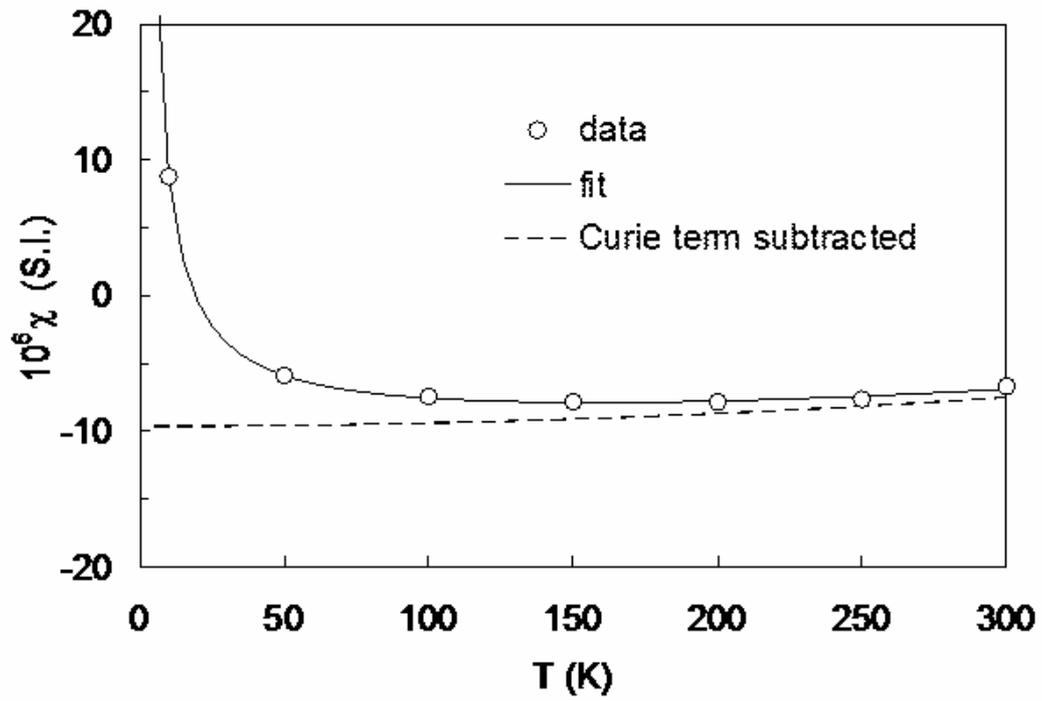

**2**

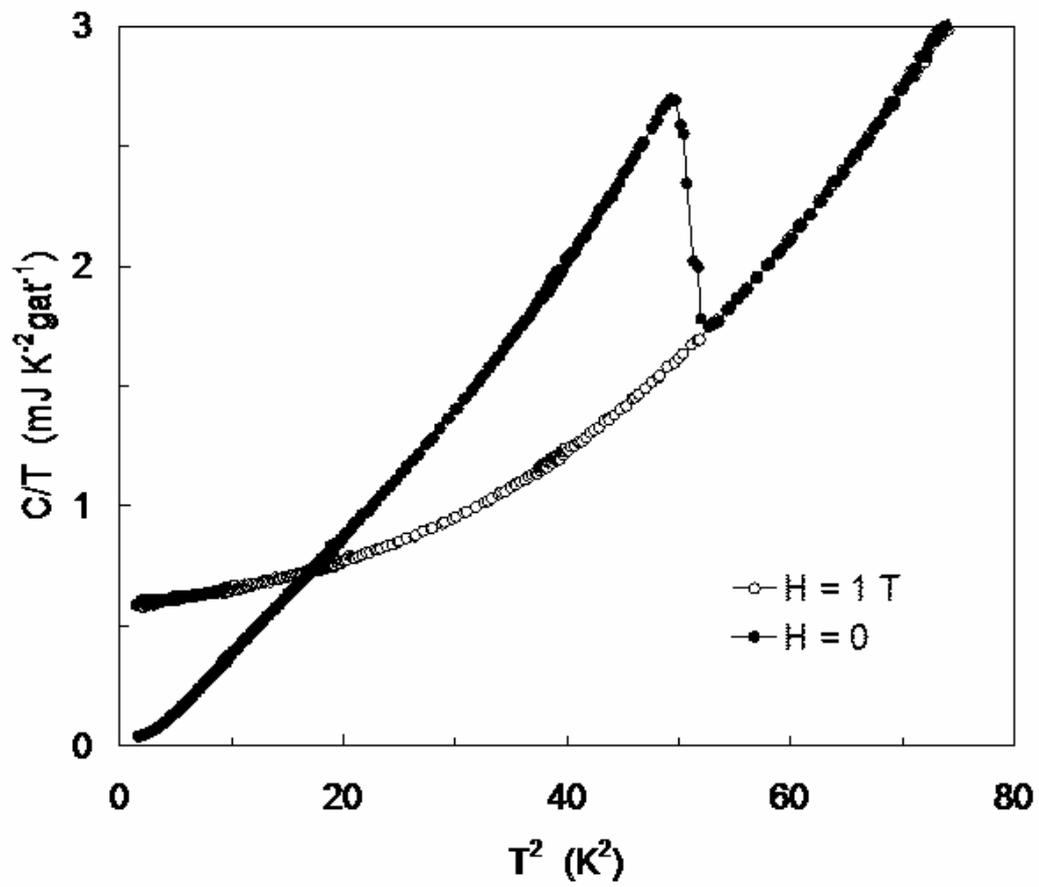

**3**

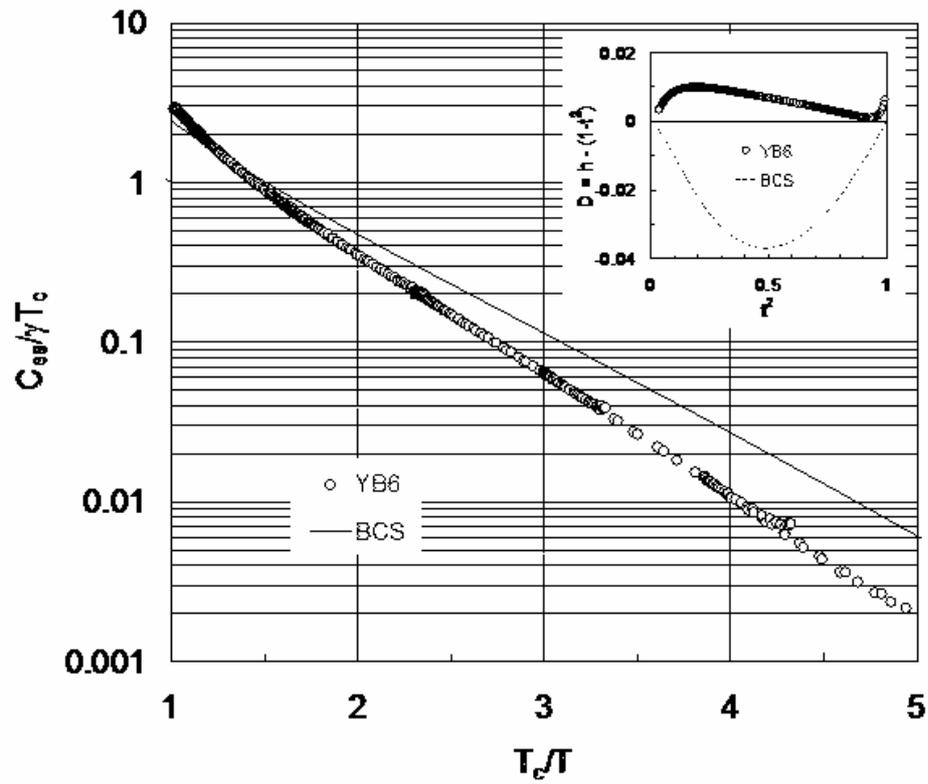

**4**

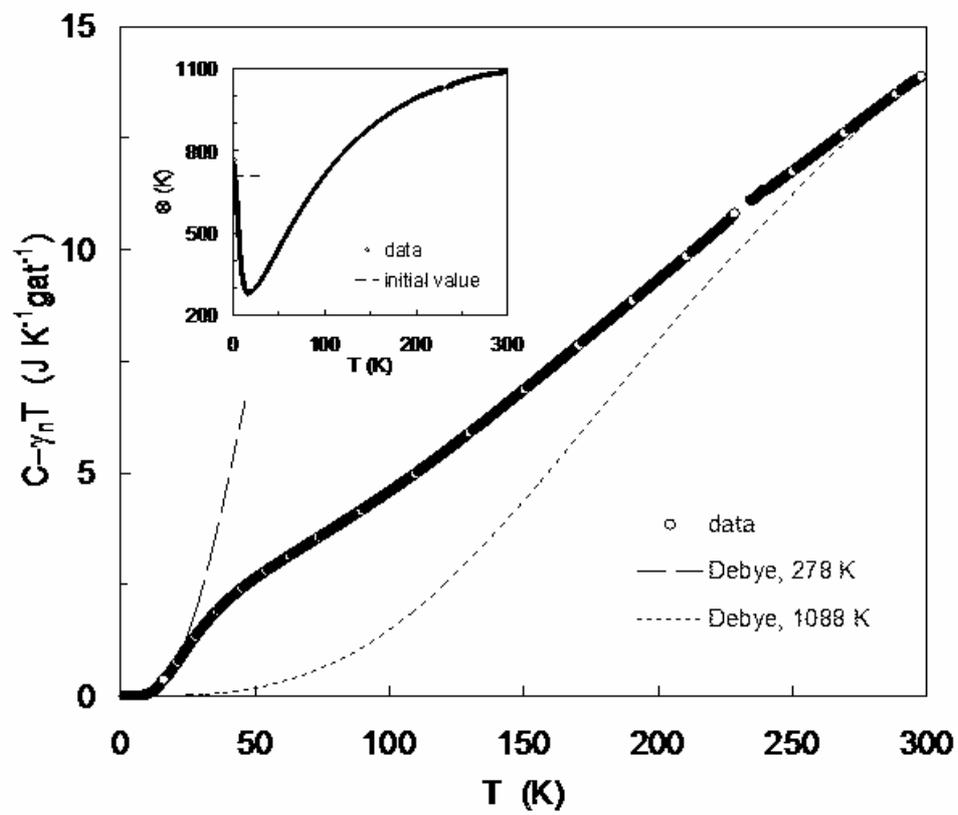

**5**

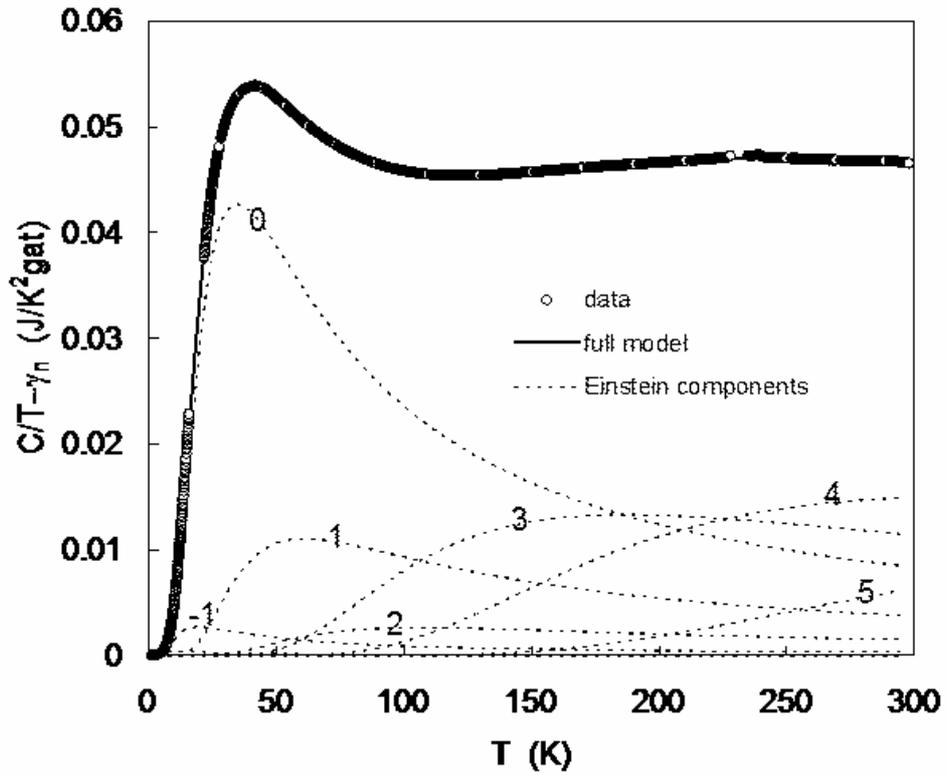

**6**

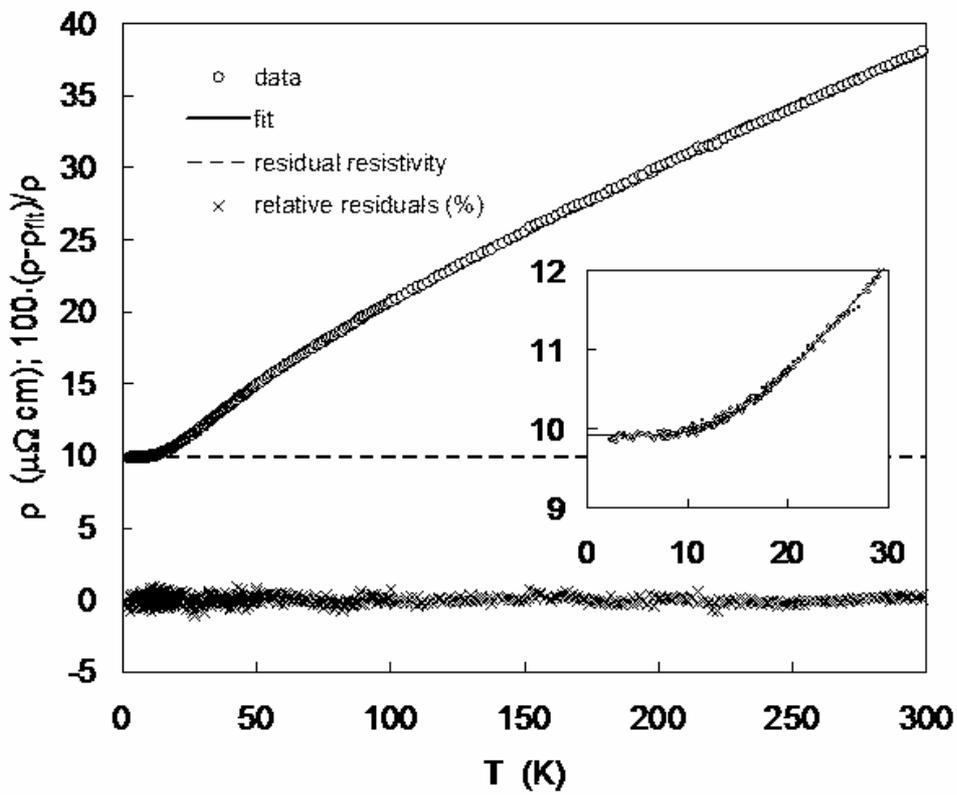

**7**

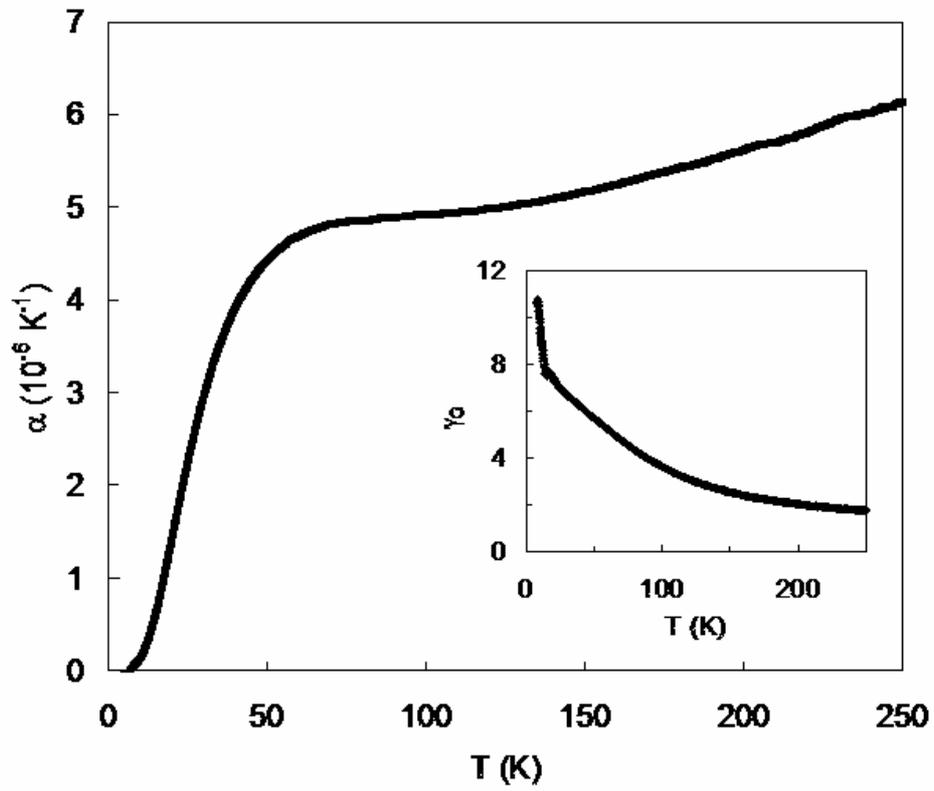

**8**

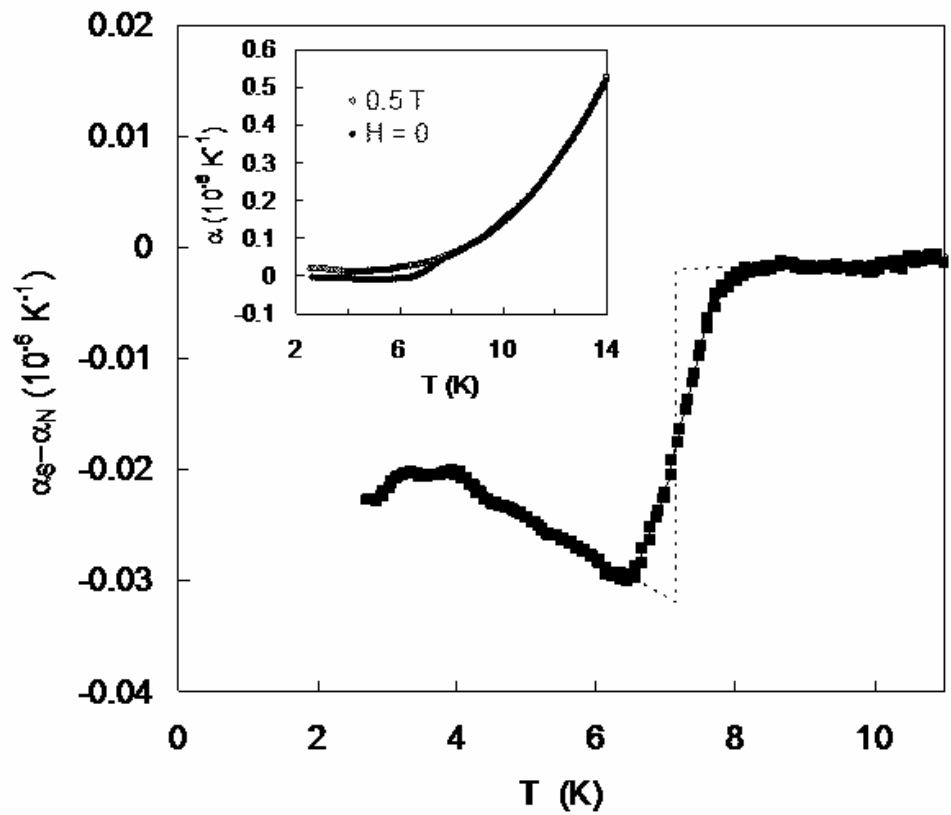

**9**

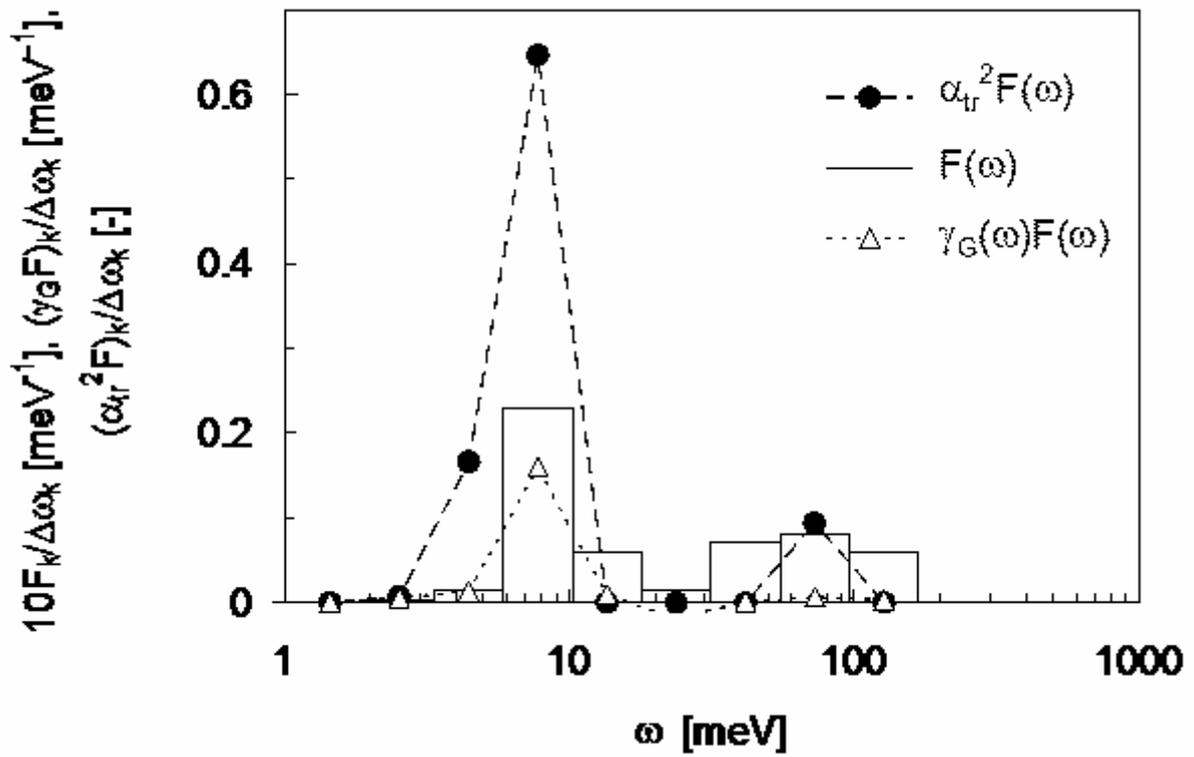

**10**

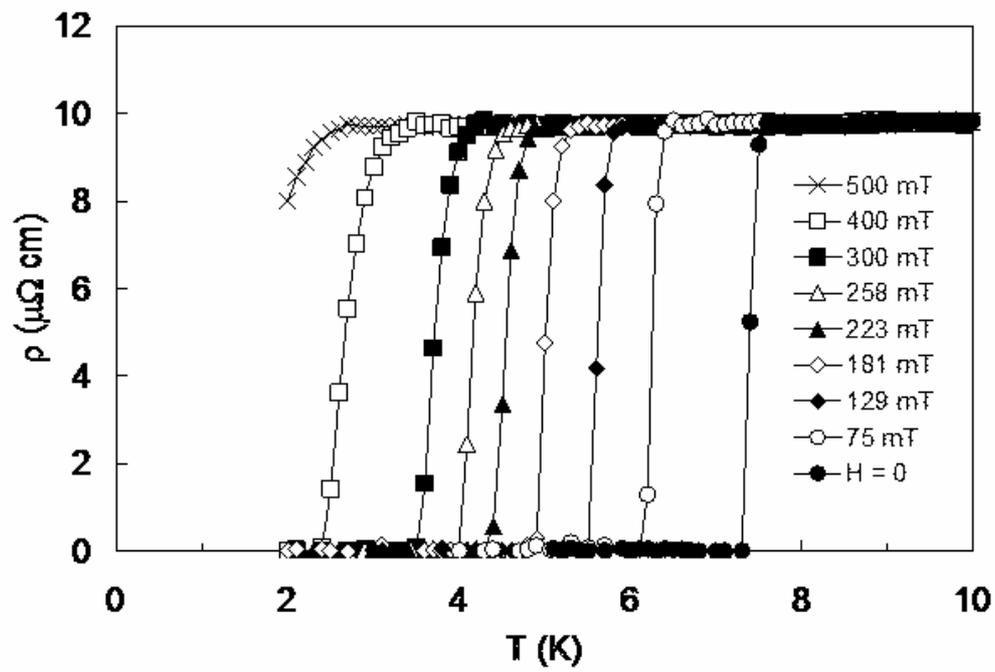

**11**

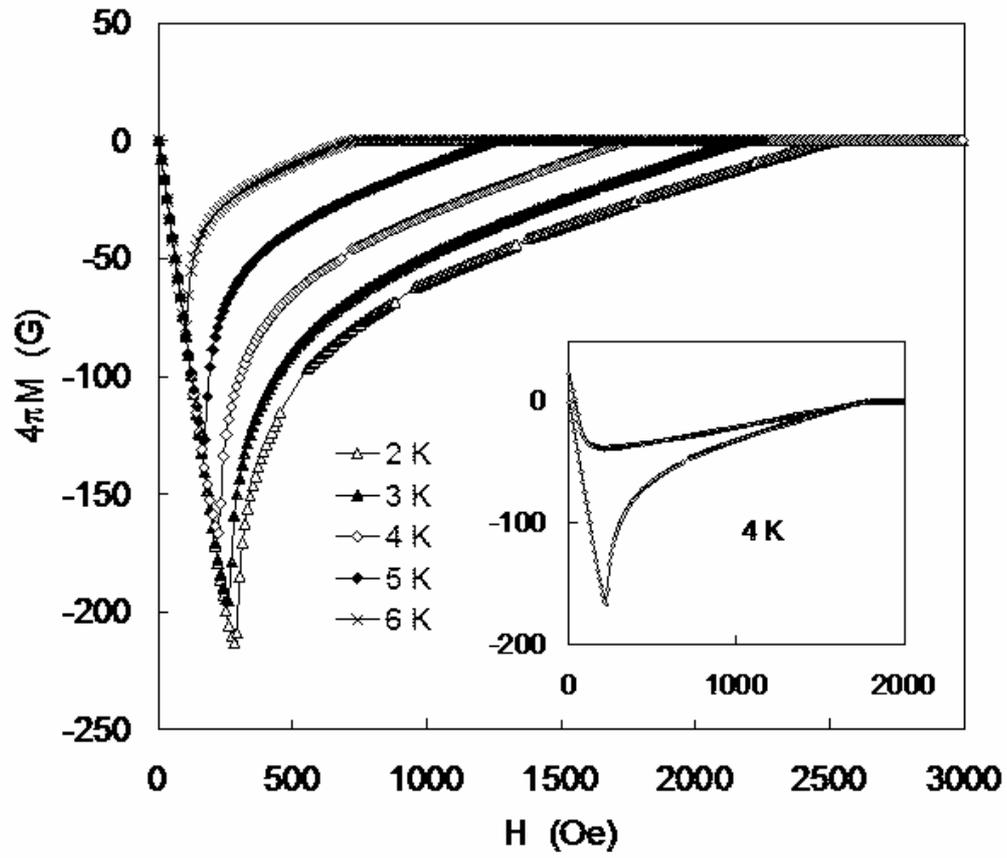

**12**

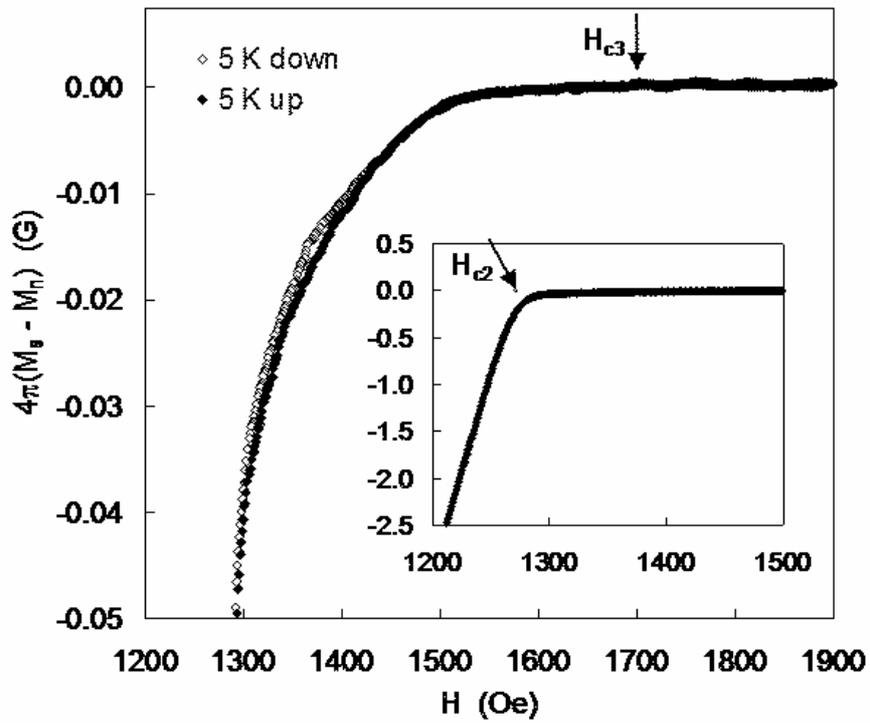

**13**

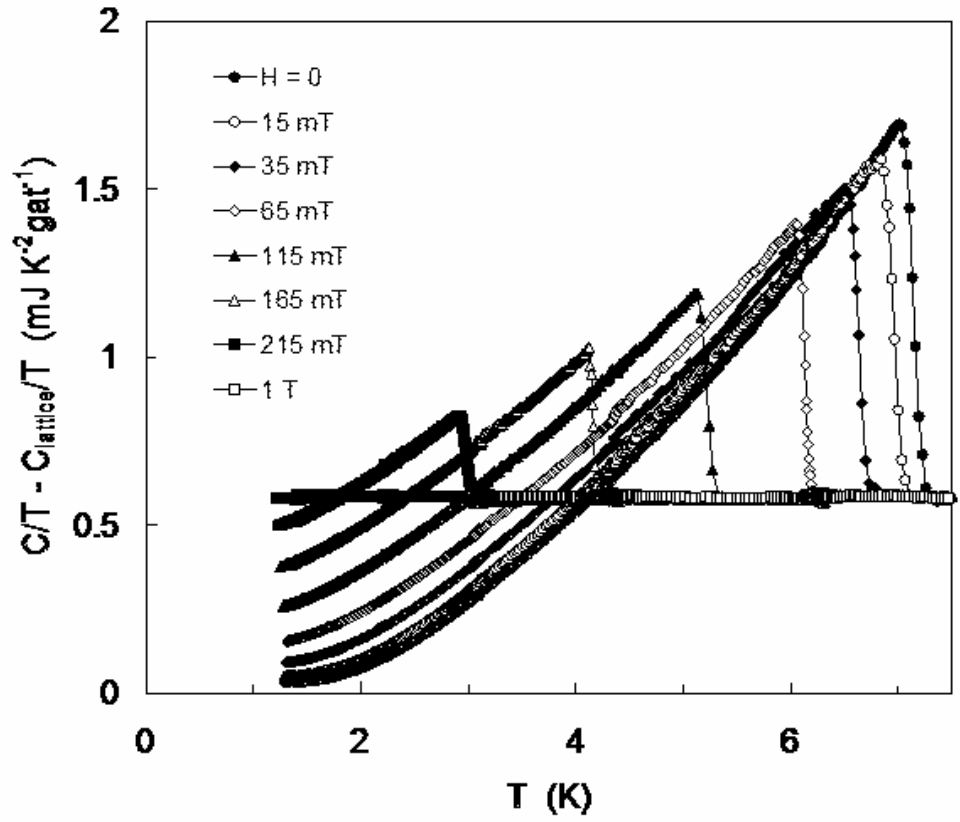

**14**

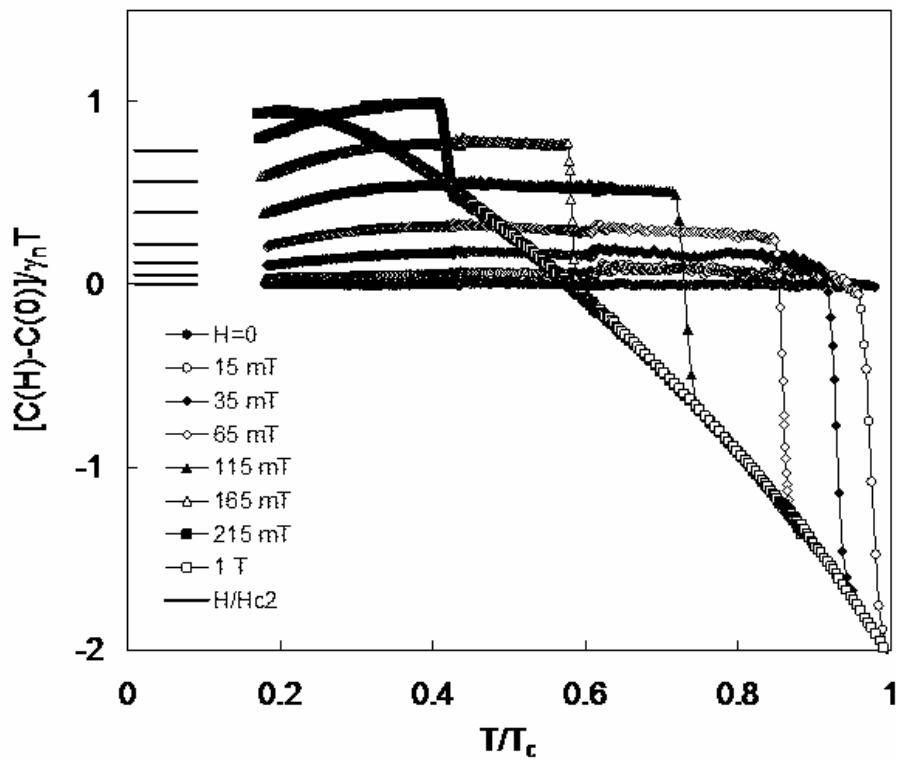

**15**

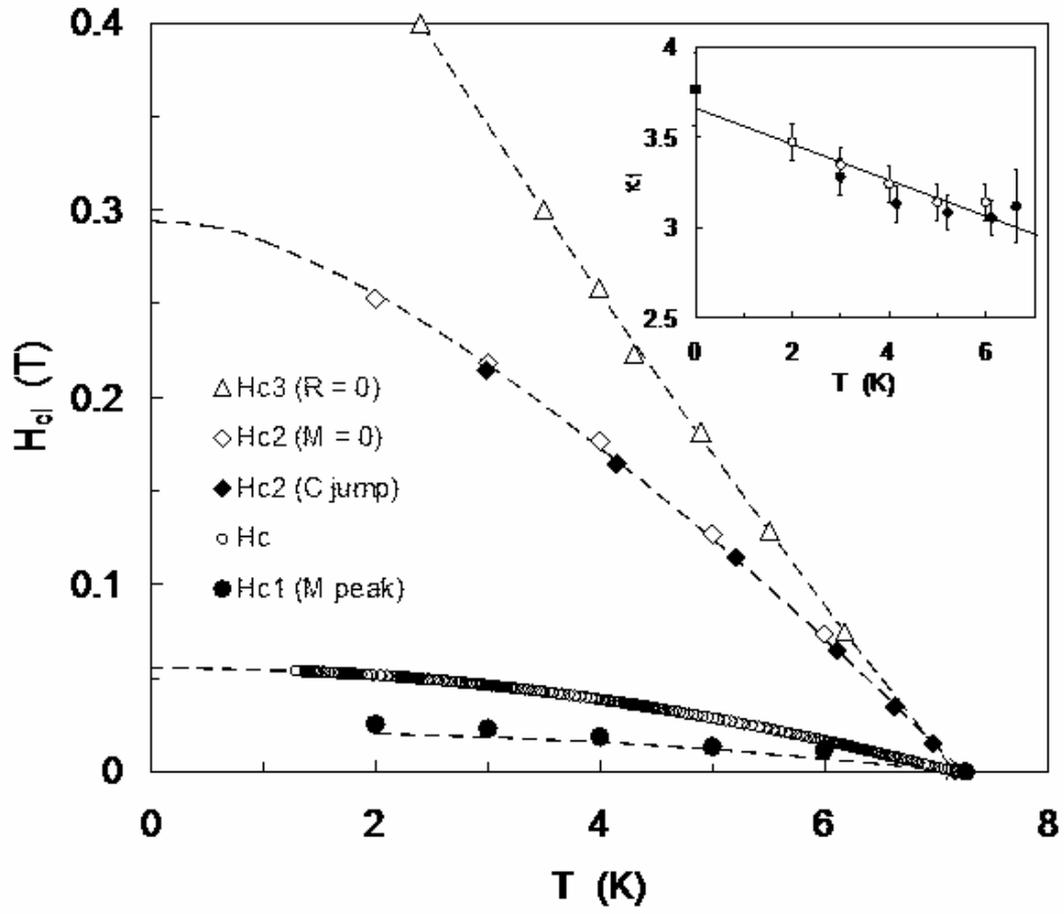

**16**

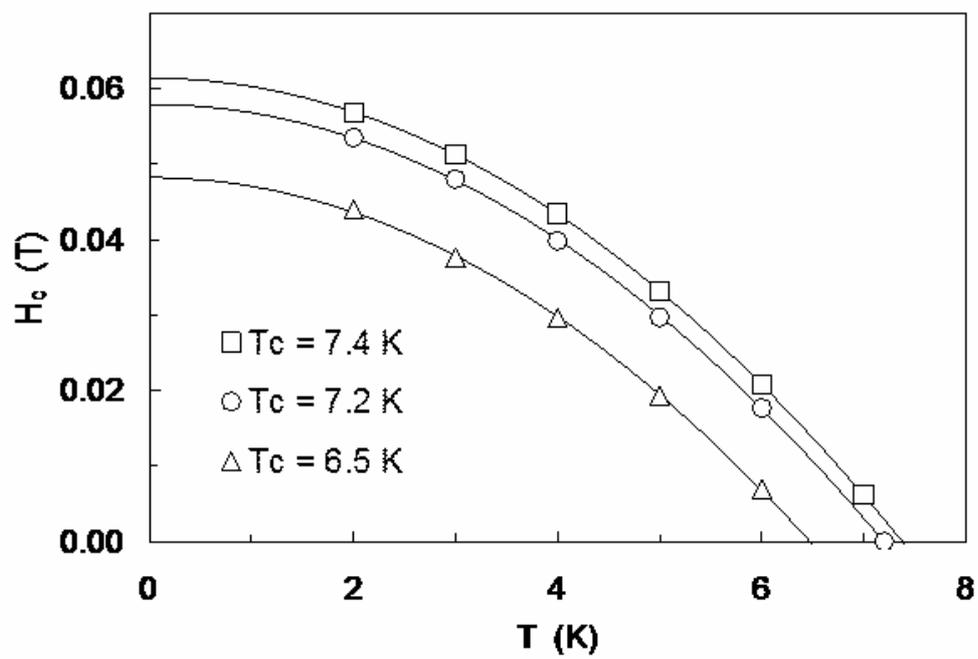

**17**